\documentclass[preprint]{vgtc}             % electronic version

\ifpdf%                                % if we use pdflatex
  \pdfoutput=1\relax                   % create PDFs from pdfLaTeX
  \usepackage{graphicx}                % allow us to embed graphics files
  \DeclareGraphicsExtensions{.pdf,.png,.jpg,.jpeg} % for pdflatex we expect .pdf, .png, or .jpg files
\else%                                 % else we use pure latex
  \usepackage{graphicx}                % allow us to embed graphics files
  \DeclareGraphicsExtensions{.eps}     % for pure latex we expect eps files
\fi%

\graphicspath{{figures/}{pictures/}{images/}{./}} % where to search for the images

\usepackage{microtype}                 % use micro-typography (slightly more compact, better to read)
\PassOptionsToPackage{warn}{textcomp}  % to address font issues with \textrightarrow
\usepackage{textcomp}                  % use better special symbols
\usepackage{mathptmx}                  % use matching math font
\usepackage{times}                     % we use Times as the main font
\usepackage{cite}                      % needed to automatically sort the references
\usepackage{tabu}                      % only used for the table example
\usepackage{booktabs}                  % only used for the table example
\onlineid{0}

\vgtccategory{Research}

\vgtcinsertpkg

\title{The Influence of Extended Reality and Virtual Characters' Embodiment Levels on User Experience in Well-Being Activities}

\author{Tanja Kojić$^1$\thanks{e-mail: tanja.kojic@tu-berlin.de} 
\and Maurizio Vergari$^1$ \thanks{e-mail: maurizio.vergari@tu-berlin.de}
\and Marco Podratz$^1$ \thanks{e-mail: marco.podratz@campus.tu-berlin.de}
\and Sebastian Möller$^{1,2}$ \thanks{e-mail: sebastian.moeller@tu-berlin.de}
\and Jan-Niklas Voigt-Antons$^3$\thanks{e-mail: jan-niklas.voigt-antons@hshl.de}}
\affiliation{\scriptsize $^1$Quality and Usability Lab, TU Berlin, $^2$German Research Center for Artificial Intelligence (DFKI)\\ $^3$Immersive Reality Lab, Hamm-Lippstadt University of Applied Sciences}

\abstract{
Millions of people have seen their daily habits transform, reducing physical activity and leading to mental health issues. This study explores how virtual characters impact motivation for well-being. Three prototypes with cartoon, robotic, and human-like avatars were tested by 22 participants. Results show that animated virtual avatars, especially with extended reality, boost motivation, enhance comprehension of activities, and heighten presence. Multiple output modalities, like audio and text, with character animations, improve the user experience. Notably, the cartoon-like character evoked positive responses. This research highlights virtual characters' potential to engage individuals in daily well-being activities.
} % end of abstract

\CCScatlist{
  \CCScatTwelve{Human-centered computing}{Visu\-al\-iza\-tion}{Visu\-al\-iza\-tion techniques}{};
  \CCScatTwelve{Human-centered computing}{Visu\-al\-iza\-tion}{Visualization design and evaluation methods}{}
}

\begin{document}


\firstsection{Introduction}

\maketitle

%% \section{Introduction} %for journal use above \firstsection{..} instead
Well-being is a comprehensive concept that includes both emotional contentment and optimal functioning \cite{wellbeingJackson}. Unfortunately, the recent onset of the COVID-19 pandemic has triggered a decline in well-being, with detrimental effects on physical and mental health. The World Health Organization (WHO) prescribes daily moderate physical activity as a means to enhance well-being, yet a significant portion of the population fails to adhere to these guidelines \cite{hildebrandt2004trendrapport}.
A recent study estimates that an impressive 110,000 deaths in the United States alone could be averted through the promotion of moderate to vigorous physical activity \cite{deaths_prevented}.

The visual presentation of characters holds significant importance in the design of virtual character systems, as it has the potential to evoke a wide range of emotions in users, consequently exerting a substantial impact on the overall user experience (UX). A prominent illustration of this phenomenon is the uncanny valley, wherein individuals often experience a sense of aversion when engaging with or observing humanoid robots that closely resemble humans.

Previous work usually assumed a consistent emotional response to virtual characters across different user groups. However, recent studies have revealed that emotional reactions can vary significantly among diverse user populations. In a study, participants encountered virtual characters with distinct appearances and motions in Virtual Reality. The research confirmed that appearance and motion significantly affected emotional responses, with females exhibiting generally more negative reactions to specific characters \cite{zombieMaleMotion}.
In a separate study, male participants embodied by muscular avatars in a Virtual Reality setting demonstrated superior performance in physical exercises compared to those with non-muscular avatars. This effect was specific to male participants, with no corresponding correlation observed among females \cite{flex}. 
These studies highlight the importance of character appearance in terms of user motivation and performance. 

Existing research often overlooks the potential of virtual characters in well-being exercises for a broad user base. While studies have explored their application among fitness enthusiasts during dedicated workouts, little to no research addresses a flexible approach for non-exercisers. This gap presents an opportunity to create a system and virtual character suited for users seeking convenient, low-commitment exercises. The aim of the project was to rebuild the "Well-being Hacks" as an interactive experience guided by virtual characters and to determine the impact of virtual characters on user motivation and engagement.

\section{Experiment Design}
%Test set-up
The experiment used a combination of qualitative and quantitative techniques to assess users' experiences with physical activity prototypes. Demographic information, sports motivation, and mood were collected from the participants. To measure the level of presence, flow, and emotions during prototype use, the Flow State Scale (FSS) \cite{jackson1996development}, Adapted Igroup Presence Questionnaire (IPQ) \cite{IPQwebsite}, and Pick a Mood \cite{desmet2012pick} were completed. In addition, after each exercise, participants provided verbal feedback in which they shared their thoughts and experiences with the prototypes. Finally, participants were asked to rate their preferred prototype, virtual character, and exercise, as well as explain why they chose them.

\begin{figure}[h!]
\centering
\includegraphics[width=0.5\textwidth]{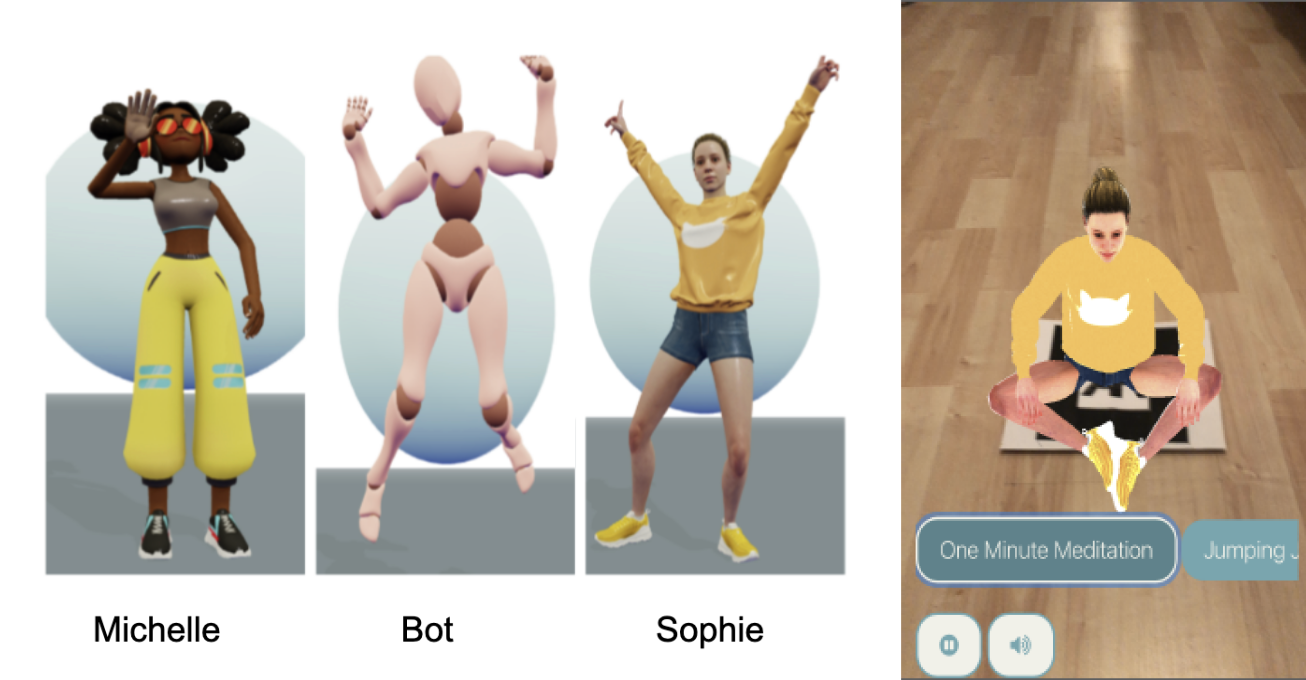}
\caption{Virtual Characters with diverse appearances from robot-like to human-like; and an example of mobile prototype with interactive exercise list and meditation exercise.}
\label{fig:char}
\vspace{0em}
\end{figure}

%Protype set-up
We opted to develop three high-fidelity prototypes, each featuring a virtual coach character. 
Within these prototypes, users had the opportunity to interact with virtual characters of varying appearances, including human-like (Sophie), cartoon-like (Michelle), and robot-like (Bot), as illustrated in Figure \ref{fig:char}. These 3D characters were open-source and readily available for download from Mixamo.com. They came equipped with predefined animations, categorized as social animations to enhance the environment's appeal and practical animations to demonstrate proper exercise techniques. Practical animations were fully crafted using Blender Studio and incorporated in Figma's final prototype.
The prototypes included instructions for four exercises: cross-crawling, jumping jacks, one-minute meditation, and side stretches. Each exercise provided various health benefits, such as stress reduction, improved physical and mental well-being, and increased flexibility \cite{smartLearning}.
Three prototypes use different technologies, output modalities, and different combinations of virtual characters: The Static Prototype with step-by-step text and static images of the character's movements on PC; The Desktop Prototype with step-by-step audio and step-by-step text fully animated characters on PC; The XR Prototype with step by step audio and fully animated characters on a smartphone. 

%procedure
The study involved a total of 36 conditions, which resulted from the combination of three different prototypes, three virtual characters, and four distinct well-being exercises. 
These conditions were allocated to participants using a Latin square randomization method, with additional considerations given to factors such as the participant's age, time of day, and physical condition.
%For instance, participants arriving after lunch were not assigned to perform activities like Jumping Jacks, recognizing the potential discomfort associated. Careful attention was also given to the overall distribution of exercises to ensure a balanced experience for each participant. %preventing any exercise from being overlooked.

\section{Results}
Twenty-two participants took part in the experiment, comprising 13 university students and 9 individuals primarily working in software development. On average, participants had a mean age of 31, with males averaging 33 years and females averaging 28 years. 

Diverse analysis techniques and tools were employed to process the collected data. Statistical significance of the overall QoE, FSS, and Adapted IPQ questionnaire was assessed using a one-way analysis of variance (ANOVA) conducted with SPSS.
Additionally, qualitative feedback from the final questions was analyzed using the Affinity Diagramming technique 

%All
\begin{table}[htbp]
\centering
  \caption{Descriptives for Overall Quality of Experience (QoE), Flow (FSS) and Presence (IPQ) over three prototypes (The Static, The Desktop and The XR Prototype). Significance is marked by asterisk. }
    \begin{tabular}{llclcll}
    \toprule
\multicolumn{1}{l}{}&  \multicolumn{2}{c}{QoE}&  \multicolumn{2}{c}{FSS}& \multicolumn{2}{c}{IPQ}\\
 Type& Mean&SD& Mean&SD& Mean& SD\\
    \midrule
         Static  &  3.50*&  1.92 &  3.38*&  .83 & 2.02*,**& .44 \\
         Desktop  &  5.41*,**&  1.05 &  3.93*&  .60 & 2.57*& .53 \\
         XR  &  4.77**&  1.60 &  3.62 &  .50 & 2.75**& .68 \\
         Total &  4.56 &  1.74 &  3.64 &  .69 & 2.45 & .63 \\
    \bottomrule
    \end{tabular}
  \label{tab:Descriptives-OverallQualityofExperience}%
\end{table}%

%QoE 
The one-way ANOVA revealed that there was a statistically significant difference in the mean of perceived overall quality score (F(2, 63) = [8.467], p = 0.001).
Tukey’s HSD Test for multiple comparisons found that the mean value of perceived overall quality was significantly different between the Static prototype and both the Desktop Prototype  (p = 0.001, 95\% C.I. = [-3.04, -0.78]) and the XR prototype  (p = 0.024, 95\% C.I. = [-2.41, -0.14]).

%Flow
Furthermore, when it comes to the feeling of flow, a statistically significant difference was found in the mean of the feeling of flow between at least two groups (F(2, 63) = [3.943], p = 0.024). Again, a multiple comparisons found that the mean value of flow was significantly different between the Static prototype and the Desktop prototype  (p = 0.018, 95\% C.I. = [-1.03, -0.79]). 

%Presence
Similar trend has been found as well when it comes to reported feeling of presence from participants. There is statistically significant difference in the mean of the perceived feeling of presence (F(2, 63) = [3.179], p = <0.001).
In particular, the mean value of the feeling of presence was significantly different both between the Static prototype and the Desktop prototype (p = 0.006, 95\% C.I. = [-0.95, -0.14]) and between the Static prototype and the XR prototype (p = 0.001, 95\% C.I. = [-1.14,-0.33]) 

%Virtual Character Preference
Participants were also asked to state their preferred virtual character and prototype preference. Participants (13 of them) favoured Michelle as their preferred virtual character due to her 'fun' and 'cool' appearance, with some desiring a sportier outfit, while Sophie was chosen (by 8 of them) for her perceived 'realistic' look.
%Prototype Preference
Regarding prototype preference, the Desktop option was favored by 13 participants for its motivation, immersion, and practicality, while the XR prototype was chosen by 9. No votes were cast for the Static prototype.
%Regarding prototype preference, the Desktop prototype got the most support with 13 participants, followed by the XR prototype, chosen by 9 participants, while the Static prototype received no votes. Participants favoured the Desktop prototype for its motivating and immersive qualities, as well as its practical usability due to screen size, timer, multiple outputs (text, character movements, audio), and background music.
%Motivation
Finally, of the 22 participants, 20 intended to use their preferred prototype and character combo, citing increased exercise motivation. Other reasons included companionship from the chosen character, the challenge it presented, and for Desktop fans, features like the timer, exercise benefits, and music that enhanced motivation.

%of the 22 participants, 20 expressed their intention to use their preferred prototype and character combination, citing increased motivation for practising the exercises. Other reasons mentioned by participants included the sense of companionship provided by the preferred character, the challenge it presented, and, for those favouring the Desktop prototype, the importance of features like the timer, exercise benefits, and music in enhancing motivation.

%Motivation

\section{Discussion \& Conclusion}
This study assessed the impact of virtual character appearance and embodiment on user experience in a well-being system. Findings revealed that animated, cartoon-like characters had a more positive influence on user experience and emotional well-being than bot and human-like characters. The Desktop and XR prototypes, featuring animations, provided a superior user experience compared to the static prototype.

We explored virtual character integration into well-being exercises using adaptable daily-use prototypes. Participants demonstrated a strong understanding of the exercises, with only 7 minor errors out of 60. The Desktop prototype excelled, offering various output modalities, a larger screen, and a visual timer. On the other hand, the XR prototype's character integration did not significantly enhance the sense of flow, while the static prototype, lacking animations, music, and voice, underperformed. Participants desired audio instructions and background music in the Static prototype, emphasizing the importance of user control over audio settings. Sophie was preferred in the XR environment, potentially due to her perceived realism, which was noteworthy. Participants also desired exercise customization options, such as duration and movement speed.

In conclusion, animated, cartoon-like characters show promise in enhancing user engagement and motivation in well-being exercises. Most participants favored their selected character and prototype combination, indicating its potential to boost motivation for well-being activities. These findings underscore the significance of valued, animated virtual characters in promoting user engagement.

%\acknowledgments{
%The authors wish to thank A, B, and C. This work was supported in part by a grant from XYZ.}

%\bibliographystyle{abbrv}
\bibliographystyle{abbrv-doi}

\bibliography{template}
\end{document}